\documentclass[sn-mathphys]{sn-jnl}
\jyear{2026}
\theoremstyle{thmstyleone}

\theoremstyle{thmstyletwo}

\theoremstyle{thmstylethree}

\begin{document}

\title[Sterile Neutrinos at MAPP in the $B-L$ Model]{Sterile Neutrinos at MAPP in the $B-L$ Model}
\author*[1]{\fnm{Frank F.} \sur{Deppisch}}\email{f.deppisch@ucl.ac.uk}
\author[2]{\fnm{Suchita} \sur{Kulkarni}}\email{suchita.kulkarni@uni-graz.at}
\author[3]{\fnm{Wei} \sur{Liu}}\email{wei.liu@njust.edu.cn}

\affil*[1]{\orgdiv{Department of Physics and Astronomy}, \orgname{University College London}, \orgaddress{\street{Gower Street}, \city{London}, \postcode{WC1E 6BT}, \country{United Kingdom}}}

\affil[2]{\orgdiv{Institute of Physics, NAWI Graz}, \orgname{University of Graz}, \orgaddress{\street{Universit\"atsplatz 5}, \city{Graz}, \postcode{8010}, \country{Austria}}}

\affil[3]{\orgdiv{Department of Applied Physics}, \orgname{Nanjing University of Science and Technology}, \orgaddress{\city{Nanjing}, \postcode{210094}, \country{P. R. China}}}

\abstract{The possibility of searching for right-handed neutrinos at the MoEDAL's Apparatus for Penetrating Particles (MAPP) detector is investigated in this work. In particular, pair-production of right-handed (RH) neutrinos $N$ from either a $B-L$ gauge boson $Z'$, as well as Standard Model (SM) $Z$ boson are considered. Under a no-background assumption, we show that the MAPP detector can be sensitive to active-sterile neutrino mixing strengths as low as $V_{\mu N}^2 \approx 10^{-12}$ for multiple choices of $m_N / m_{Z'}$ values, when taking the $B-L$ gauge coupling $g_{B-L} = 10^{-3}$ near its current limit. The SM $Z$ boson portal can reach a similar sensitivity, when the effective mixing between the  $B-L$ and SM gauge boson is $\alpha \approx 0.002$.}


\maketitle

\section{Introduction}
The lightness of active neutrinos is a strong indication that the mass generation mechanism responsible is different from that of the other fermions, and is based on physics beyond the Standard Model (SM). A simple, ultraviolet-complete model based on the additional gauge group $U(1)_{B-L}$~\cite{Davidson:1978pm, Mohapatra:1980qe} (baryon minus lepton number) can be employed which incorporates a seesaw mechanism with three right-handed Majorana neutrinos. It has a SM-singlet Higgs field $\chi$ which acquires a vacuum expectation value that spontaneously breaks the $B-L$ symmetry leading to a massive $Z'$ boson and heavy right-handed (RH) Majorana neutrinos $N_i$. 
	
The $Z'$ gauge boson decays into pairs of RH neutrinos providing an efficient means of their production, even if the active-sterile mixing $V_{lN}$ between the mostly sterile (under SM interactions) RH neutrinos and the active neutrinos is small. In this case, distinctive displaced vertex signatures at the Large Hadron Collider (LHC) are produced as the RH neutrino must decay via the small active-sterile mixing corresponding to long lifetimes. This production process $pp\to Z' \to NN$ at a hadron collider is controlled by the $B-L$ coupling  $g_{B-L}$, the mass $m_{Z'}$ of the heavy gauge boson and the masses $m_{N_i}$ of the heavy neutrinos. 

The primary exotic particle, the heavy gauge boson, can be searched for in different ways. LHC searches for a heavy resonance in dilepton final states put a bound $m_{Z'} > 4.5$~TeV~\cite{Aaboud:2017buh} for a $B-L$ gauge coupling equal to that of the SM $Z$ boson. While the $B-L$ breaking scale $\Lambda_{B-L} = M_{Z'} / (2g_{B-L})$ is constrained to be larger than 3.45~TeV from LEP-II \cite{Heeck:2014zfa, Cacciapaglia:2006pk, Anthony:2003ub, LEP:2003aa, Carena:2004xs}, these limits fail when $m_{Z'}$ becomes too small as they assume an effective interaction with the $Z'$ being integrated out. Limits on the full $(m_{Z'},  g_{B-L})$ parameter space can be derived using the Constraints On New Theories Using Rivet (CONTUR) tool for $g_{B-L} < 10^{-3}$ approximately for a $Z'$ above 1~GeV~\cite{Amrith:2018yfb, Butterworth:2016sqg}. Dark photon searches in various probes can be recast using the Darkcast tool~\cite{Ilten:2018crw}, see Figs.~\ref{fig:userzp} and \ref{fig:userzpn}. $g_{B-L}$ can also be constrained by the violent stellar explosions such as gamma ray bursts from pair annihilation of neutrinos, with less stringent limits~\cite{Poddar:2022svz, A:2023wup}.

Limits on the heavy RH neutrino sector can be derived from sterile neutrino searches, which are based on the SM production and decay mechanisms, see, e.g., Refs.~\cite{Bolton:2019pcu, Abdullahi:2022jlv, Bolton:2022pyf} and references therein. As mentioned, the RH neutrinos are long-lived if the active-sterile mixing $V_{lN}$ is small which is generically expected in a canonical Type-I seesaw mechanism of neutrino mass generation. Here, the mass scale $m_\nu$ of light neutrinos is related to that of the RH neutrinos as $m_\nu \sim V_{lN}^2 m_N$. With sub-eV light neutrinos, as required by $0\nu\beta\beta$ and Tritium beta decay experiments \cite{Kraus:2004zw, Aseev:2011dq, Deppisch:2016rox} as well as cosmological observations \cite{Ade:2015xua}, the active-sterile neutrino mixing is very small, $V_{lN}^2 \approx 10^{-10} - 10^{-12}$, for $m_N \approx 10 - 100$~GeV. The RH neutrino decay lengths are then of the order of centimetres to metres and models of neutrino mass generation around the electroweak scale thus clearly motivate searches at the lifetime frontier. We should mention however that this is not necessarily the case in extended scenarios such as the inverse seesaw mechanism where the active-sterile neutrino mixing is large but the RH neutrinos are quasi-Dirac states, see, e.g., Ref.~\cite{Deppisch:2015qwa} and references therein.
	
Searches for displaced vertices at the LHC and in dedicated experiments have gained significant attention, see, e.g., Refs.~\cite{Feng:2022inv, MammenAbraham:2020hex, Alimena:2019zri}. Several CMS and ATLAS analyses \cite{CMS:2014hka, CMS:2015pca, CMS:2014mca} have studied displaced vertices of long-lived neutral particles in various mass ranges, with no definite signal observed so far. Searches for heavy sterile neutrinos, including displaced vertex signatures, have been studied in Refs.~\cite{Khachatryan:2015gha, Sirunyan:2018mtv, Aad:2019kiz}.
	
We here focus on the signal produced by the process $pp \to NN$, specifically in the MoEDAL’s Apparatus for Penetrating Particles (MAPP) displaced vertex detector \cite{Acharya:2022nik}. We regard this distinct signal as background free for large decay lengths $\approx 10$~m. Even under the highly constrained parameter space, the channel $pp\to Z' \to NN$ is promising to search for with MAPP. Phase-1 of the experiment has received approval from the CERN council for the LHC Run-3. This is also the case for the the ForwArd Search ExpeRiment (FASER)~\cite{Ariga:2018uku} whereas other proposed detectors of a similar type include the MAssive Timing Hodoscope for Ultra Stable neutraL
pArticles (MATHUSLA)~\cite{Chou:2016lxi} and the COmpact Detector for EXotics at LHCb (CODEX-b)~\cite{Gligorov:2017nwh}. This paper is an extension of \cite{Deppisch:2019kvs} where we update our results for the expanded MAPP detector geometry of Phase-2 \cite{Acharya:2022nik}, in comparison with the detectors FASER, MATHUSLA and CODEX-b. We also expand the $B-L$ model to include a kinetic mixing between the SM $Z$ and $B-L$ $Z'$ which triggers the process $pp \to Z \to NN$.
	
Similar studies have been made to search for RH neutrinos. Ref.~\cite{Accomando:2017qcs} discussed the same processes $pp\to Z' \to NN$ focussing on prompt, short-lived RH neutrinos. Ref.~\cite{Batell:2016zod} studied displaced vertices at the LHC and demonstrated that the proposed SHiP~\cite{Alekhin:2015byh, Anelli:2015pba} detector yields a competitive probe to constrain the active-sterile mixing. A displaced vertex search at LHCb has been proposed for boosted light RH neutrinos~\cite{Antusch:2017hhu}. Ref.~\cite{Deppisch:2018eth} shows the potential to probe RH neutrinos with displaced vertices at the current LHC run, the high luminosity run and at proposed lepton colliders with a limit $V_{\mu N}^2 < 10^{-14}$. The production of the $B-L$ gauge boson $Z'$ at lepton colliders has also been studied \cite{Ramirez-Sanchez:2016ugz}. Several analyses have focused on the seesaw aspect of the $B-L$ model. For example, the gauge $B-L$ scenario may be applied to a low scale seesaw mechanism \cite{Khalil:2006yi}, an inverse seesaw scenario \cite{Khalil:2010iu} and linear seesaw scenario \cite{Dib:2014fua}. Different aspects of the $B-L$ model have been studied in \cite{Das:2017flq, Das:2017deo, Das:2018tbd, Jana:2018rdf}. 
	
The paper is organized as follows. In Section~\ref{blreview}, we briefly review the $B-L$ model, including a potential mixing between the SM $Z$ and $B-L$ $Z'$, the process $pp \to Z' \to NN$ and current limits on the $B-L$ model. We then discuss experimental considerations, namely the signal trigger and the detector geometry in Section~\ref{sec:detectors}. Our results, i.e., the derived sensitivity of the MAPP experiment and similar detectors to the $B-L$ model, are presented in Section~\ref{sec:simu}. We conclude our work in Section~\ref{conclu}.
	
	
\section{$B-L$ gauge model and its phenomenology}
\label{blreview}
	
\subsection{Model setup}
The $U(1)_{B-L}$ gauge model adds an Abelian gauge field $B'_\mu$, a SM-singlet scalar field $\chi$ and three RH neutrinos $N_i$ to the SM. The full gauge group is $SU(3)_c\times SU(2)_L \times U(1)_Y \times U(1)_{B-L}$, under which the scalar and RH neutrino fields $\chi$ and $N_i$ have $B-L$ charges $Y_{B-L}(\chi) = +2$ and $Y_{B-L}(N_i) = -1$, respectively, but are otherwise singlets. The SM fermions have their canonical $Y_{B-L}$ quantum numbers. The scalar sector of the Lagrangian is given by
\begin{align}
\label{Ls}
	{\cal L} &\supset 
		(D^{\mu}H)^\dagger(D_{\mu}H) 
	  + (D^{\mu}\chi)^\dagger D_{\mu}\chi \nonumber\\
	  &- \left[m^2(H^\dagger H) + \mu^2(\chi^*\chi) + \lambda_1 (H^\dagger H)^2 
	  + \lambda_2 (\chi^*\chi)^2 + \lambda_3 (H^\dagger H)(\chi^*\chi)\right],
\end{align} 
where $H$ is the SM Higgs doublet and $D_\mu$ is the covariant derivative. The latter includes the $U(1)_{B-L}$ term $g_{B-L} Y_{B-L}$, with the extra gauge coupling $g_{B-L}$. The fermionic part of the Lagrangian adds the terms
\begin{align}	
\label{Lf}
	{\cal L} &\supset
	i\overline{\nu_{Ri}}\gamma_\mu D^\mu \nu_{Ri}
	 - y_{ij}^\nu \overline{L_i}\nu_{Rj}\tilde{H}
	 - y_{ij}^M \overline{\nu^c_{Ri}} \nu_{Rj}\chi 
	 + \text{h.c.},
\end{align} 
with a summation over the fermion generations $i, j = 1, 2, 3$ implied. The first term is the kinetic energy density of the RH gauge eigenstates $\nu_{R_i}$ and the other two terms describe the Yukawa couplings with the SM lepton doublets $L_i$, and between the RH neutrinos, respectively. Here, $\tilde{H} = i\sigma^2 H^\ast$ and the Yukawa matrices $y^\nu$ and $y^M$ are a priori arbitrary. The RH neutrino masses are generated by the breaking of the $B-L$ gauge symmetry when $\chi$ acquires its vacuum expectation value $\langle\chi\rangle$. 

\subsection{Active-sterile neutrino mixing}
The mixing between the active and RH neutrinos is induced by the Dirac mass matrix $m_D = y^\nu v/\sqrt{2}$, with the SM Higgs vacuum expectation value $v$. The full $6\times 6$ mass matrix in the $(\nu_L, \nu_R^c)$ basis is
\begin{align}
\label{MD}
	{\cal M} = 
	\begin{pmatrix}
		0   & m_D \\
		m_D & M_R
	\end{pmatrix},
\end{align} 
with 
\begin{align}
\label{MDM}
	m_D = \frac{y^\nu}{\sqrt{2}}v, \quad M_R = \sqrt{2} y^M \langle\chi\rangle. 
\end{align} 
In the canonical seesaw case, $M_R \gg m_D$, the light and heavy neutrino mass matrices are
\begin{align}
\label{seesaw}
	m_\nu \sim - m_D M^{-1}_R m^T_D, \quad m_N \sim M_R,
\end{align}
with the flavour and mass eigenstates connected by  
\begin{align}
\label{Neutrino}
	\begin{pmatrix}
		\nu_L \\ \nu_R
	\end{pmatrix} = 
	\begin{pmatrix}
		V_{LL} & V_{LR} \\
		V_{RL} & V_{RR}
	\end{pmatrix}
	\begin{pmatrix}
		\nu \\ N
	\end{pmatrix}.
\end{align} 
The above are written schematically in terms of $3\times 3$ block matrices and three-component field vectors. For example, the top left element is the SM charged current lepton mixing matrix $V_{LL} = U_\text{PMNS}$, apart from small corrections due to non-unitarity. Likewise, $V_{LR}$ describes the admixture of the RH neutrinos in the light states. We mainly consider the case of a single generation of light and RH neutrino where the above reduce to a $2\times 2$ form. We thus assume that generations decouple and the Yukawa coupling matrix is diagonal ($i = e, \mu, \tau$),
\begin{align}
\label{YY}
	y^\nu_{ii} = \frac{\sqrt{2} m_{N_i} V_{iN}}{v},
\end{align} 
where $V_{e N}$, $V_{\mu N}$, $V_{\tau N}$ are the active-sterile mixing per generation.
	
\subsection{Gauge mixing}
\label{sec:kineticmixing}
In our previous paper \cite{Deppisch:2019kvs}, we did not include a potential kinetic and mass mixing between the SM hypercharge and $B-L$ gauge bosons. We here incorporate these effects by including the terms~\cite{Dev:2021otb} 
\begin{align}
	{\cal L} \supset 
		- \frac{\epsilon}{2}B^{\mu\nu} B'_{\mu\nu} 
		+ m^2_{ZZ'} Z^\mu Z'_\mu,
\end{align} 
where $B^{\mu\nu}$ and $B'_{\mu\nu}$ are the field strength tensors of SM hypercharge $U(1)_{Y}$ and $U(1)_{B-L}$, respectively. The mass mixing term violates the model's gauge symmetry but may be present if the $U(1)_{B-L}$ and electroweak symmetry breaking are not independent of each other. We treat both $\epsilon$ and $m_{ZZ'}^2$ as effective, independent parameters at the electroweak scale. The mass
mixing induces a field mixing angle $\alpha$,
\begin{align}
	\tan\alpha = \frac{m^2_{ZZ'}}{m_{Z'}^2 - m_Z^2}.
\end{align}
In combination, this results in coupling of the SM-like gauge boson to the $B-L$ fermion current $J_{B-L}^\mu$ \cite{Lindner:2018kjo},
\begin{align}
		{\cal L} \supset - \sin\theta_{ZZ'} \cdot Z_\mu J_{B-L}^\mu,
\end{align}
with the overall mixing angle
\begin{align}
	\sin\theta_{ZZ'} \approx 
	\tan\alpha 
	+ \frac{\epsilon\sin\theta_W}{m_{Z'}^2/m_Z^2 - 1},		
\end{align}
valid for small $\epsilon$ and $\alpha$. Here, $\theta_W$ is the electroweak mixing angle. The SM and the $B-L$ Higgs also mix, induced by the term $\lambda_3 (H^\dagger H)(\chi^*\chi)$ in Eq.~\eqref{Ls}, but this is not relevant for our analysis.

\subsection{Right-handed neutrino production and decay}
\label{sec:prodN}
We study the displaced decays of RH neutrinos produced via a $Z'$ boson, $pp \to Z' \to NN$. Both $Z'$ and $N$ have very small widths and in the narrow width approximation we have
\begin{align}
\label{eq:processZprime}
	&\sigma(pp \to Z' \to NN \to N + \text{displaced vertex}) = \nonumber\\
	&\sigma(pp \to Z') 
	 \times \text{BR}(Z' \to NN)
	 \times \text{BR}(N \to\text{visible}).
\end{align} 
Thus, we consider a final state consisting of a single displaced vertex with charged SM particles. The partial decay width of $Z' \to NN$ (one generation) is
\begin{align}
\label{partialdecay}
	\Gamma(Z'\to N N) = \frac{1}{6}\frac{g_{B-L}^2}{4\pi} m_{Z'}
	\left(1 - \frac{4 m_N^2}{m_{Z'}^2}\right)^{3/2}.
\end{align} 
The total $Z'$ decay width can be similarly calculated by summing over all SM fermions, $Z' \to \bar f f$ and the RH neutrinos \cite{Deppisch:2019kvs}. This simple approximation does not consider hadronization of quark pairs which could be captured using the vector meson dominance assumption~\cite{Fujiwara:1984mp}.

Alternatively, we consider production of RH neutrinos through the SM $Z$ via gauge mixing,
\begin{align}
\label{eq:processZ}
	&\sigma(pp \to Z \to NN \to N + \text{displaced vertex}) = \nonumber\\
	&\sigma(pp \to Z) 
	\times \text{BR}(Z \to NN)
	\times \text{BR}(N \to\text{visible}).
\end{align}
The cross section is $\sigma(pp \to Z \to \ell^+ \ell^-) \approx 2111$~pb at the 14 TeV LHC~\cite{Ogul:2017zjd}, and 1981~pb at the 13~TeV LHC~\cite{ATLAS:2016fij}. The branching ratio of $Z$ is
\begin{align}
	\text{BR}(Z \to NN) = \frac{\Gamma(Z \to NN)}{\Gamma(Z_\text{SM}) + \Gamma(Z \to NN)},
\end{align}
with 
$\Gamma(Z_\text{SM}) \approx 2.50$~GeV~\cite{ParticleDataGroup:2018ovx} and
\begin{align}
\label{eq:decaygaugemixing}
	\Gamma(Z \to NN) = \frac{1}{6}\frac{g^2_{B-L}\sin^2\theta_{ZZ'}}{4\pi} m_Z
	\left(1-\frac{4 m_N^2}{m_Z^2}\right)^{3/2}.
\end{align}
We note that the RH neutrino production cross section is independent of the active-sterile neutrino mixing in this case as well. In addition to the gauge mixing contribution above, there is in principle also a contribution $\sigma(pp \to Z \to NN) \propto V_{lN}^4$ due to non-unitarity effects but we neglect this for the small active-sterile mixing strengths considered.

As discussed above, we assume flavour-diagonal active-sterile neutrino mixing and we concentrate on one generation of RH neutrinos mixing dominantly with muon flavour through $V_{\mu N}$. Therefore, the $N$  has three-body decays $N\to \mu^\pm q\bar{q}$ and $N\to \mu^+\mu^-\nu_\mu$ through off-shell SM $W^{\pm(*)}, Z^{(*)}$. The corresponding branching ratios have been calculated in Ref.~\cite{Deppisch:2018eth}. Approximately, the resulting decay length for $m_N \lesssim m_Z$ is 
\begin{align}
\label{lengthapproxi}
	L_N \approx 0.025~\text{m} 
	\cdot \frac{10^{-12}}{V_{\mu N}^2} 
	\cdot \left(\frac{100~\text{GeV}}{m_N}\right)^5.
\end{align} 
For such small active-sterile mixing, as expected in the canonical seesaw mechanism, e.g., $V^2_{\mu N} \approx 10^{-8}$ for $m_N = 3$~GeV, the decay length is of the order of meters, potentially detectable in a displaced vertex experiment such as MAPP.
	
\subsection{Constraints on the $B-L$ gauge coupling}
\label{sec:darkcast}

Our main focus is on the sensitivity on the active-sterile neutrino mixing strength $V_{\mu N}^2$. If this mixing is small, as expected in the canonical seesaw, the RH neutrino cannot be produced abundantly via the SM currents. We here instead rely on the $B-L$ current regulated by the associated gauge coupling $g_{B-L}$. It is thus important to understand the existing constraints. In Ref.~\cite{Deppisch:2019kvs}, we used the tool Darkcast to determine these constraints from a wide range of observables. Here, we also included the three heavy neutrinos $N$ as final states of $Z'$ decays. 
\begin{figure}[t!]
	\centering
	\includegraphics[width=0.75\textwidth]{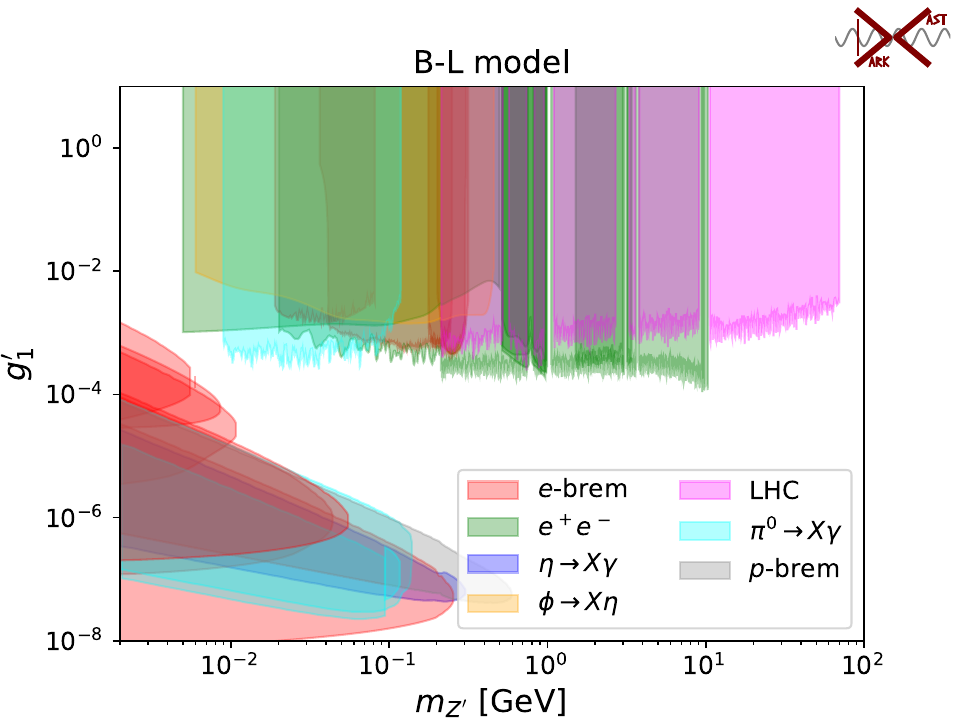}
	\caption{Constraints on the $U(1)_{B-L}$ gauge coupling $g_1' = g_{B-L}$ as a function of the $Z'$ mass $m_{Z'}$, taken from \cite{Deppisch:2019kvs} using DarkCast. The constraints are from decays to SM final states in the $B-L$ model with three generations of heavy neutrinos.}
	\label{fig:userzp}
\end{figure}
The constraints on $g_{B-L}$ are shown in Figs.~\ref{fig:userzp} and \ref{fig:userzpn} as a function of $m_{Z'}$. The constraints are from $Z'$ decays to SM particles (Fig.~\ref{fig:userzp}) and RH neutrinos (Fig~\ref{fig:userzpn}). Approximately, the constraints can be summarized as
\begin{align}
	g_{B-L} &\lesssim 10^{-4} \text{ for } 0.01~\text{GeV} \lesssim m_{Z'} \lesssim \,\,\, 10~\text{GeV}, \\
	g_{B-L} &\lesssim 10^{-3} \text{ for }\,\,\,\,\, 10~\text{GeV} \lesssim m_{Z'} \lesssim 100~\text{GeV}, 	
\end{align}
where we neglect the constraints for very small $g_{B-L}$ for $m_{Z'} \lesssim 0.1$~GeV (due to the $Z'$ being long-lived) which are not relevant for our analysis. As we will mainly focus on RH neutrino masses $m_N \gtrsim 5$~GeV but below the $Z$ mass, we use the benchmark value $g_{B-L}$ = $10^{-3}$ for $m_{Z'} > 10$~GeV and $g_{B-L}$ = $10^{-4}$ for $m_{Z'} < 10$~GeV.
\begin{figure}[t!]
	\centering
	\includegraphics[width=0.75\textwidth]{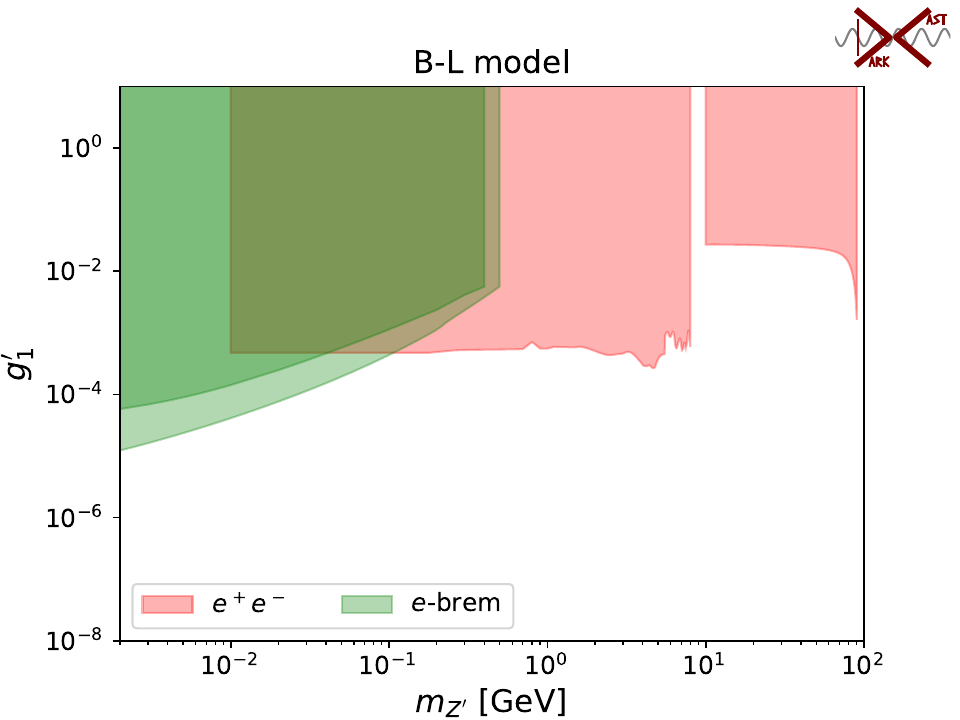}
	\caption{As Fig.~\ref{fig:userzp}, but showing constraints from decays to stable heavy neutrinos with $m_N = 0.3\times m_{Z'}$ in the $B-L$ model with three generations of heavy neutrinos.}
	\label{fig:userzpn}
\end{figure}
%

\section{Displaced vertex signal at MAPP}
\label{sec:detectors}

\subsection{Event generation}
In this section, we consider in turn the displaced decay processes in Eq.~\eqref{eq:processZprime} and Eq.~\eqref{eq:processZ}. To avoid complication from the lack of factorization and non-perturbative processes, we only simulate the case where $m_{Z'} \gtrsim 10$~GeV. We use the Universal FeynRules Output (UFO)~\cite{Degrande:2011ua} of the $B-L$ model \cite{Deppisch:2018eth} and the Monte Carlo event generator {\tt MadGraph5aMC$@$NLO} -v2.6.7~\cite{Alwall:2014hca} using the parton distribution function set nn23lo1. We create $10^5$ signal events for a given signal sample. The parton level events are passed to PYTHIA v8.235~\cite{Sjostrand:2014zea} using the default tune 14. Taking into account the proper decay length and the direction and boost of a RH neutrino in a given event, we generate $100 \times \sigma \times \mathcal{L} / 10^4$ sub-events, distributed according to the exponential decay length in the laboratory frame. The events are then selected based on the kinematic and geometric selection criteria described below. We do not perform a detailed detector simulation, nor do we consider any background events.

\subsection{MAPP detectors}
\label{sec:MAPP}
	
Within the MoEDAL experimental framework~\cite{Pinfold:2019nqj}, the MAPP detector will focus on detecting long-lived particles. Phase 1 of MAPP (MAPP-1) is currently being installed for the LHC Run-3 whereas MAPP-2 \cite{Acharya:2022nik} for the HL-LHC will considerably enhance the fiducial volume. MAPP-1 is currently being installed in the UA83 tunnel with a fiducial volume of $1.2~\text{m} \times 1.2~\text{m} \times 3~\text{m}$ whereas MAPP-2 will be located in the UGC1 up to 50~m away from the interaction point of LHCb, with 7 to 10~m depth covering 5 to 25 degrees away from the beam-line. We have simulated events for both detector geometries but due to the small size of MAPP-1, we have found no appreciable sensitivity in our model.
	
%
%
	
MAPP-2 is planned to operate at the HL-LHC with 300~fb$^{-1}$ luminosity, incorporating a larger detector volume as a planned extension for the HL-LHC, formed by two polyhedra,
\begin{align}
	\text{Point 1} &= (04.00, \phantom{-}1, -61.39), \quad
	\text{Point 2}  = (16.53, \phantom{-}1, -35.45), \nonumber\\
	\text{Point 3} &= (03.27, \phantom{-}1, -52.83), \quad
	\text{Point 4}  = (12.24, \phantom{-}1, -33.63), \nonumber\\
	\text{Point 5} &= (04.00,          - 2, -61.39), \quad
	\text{Point 6}  = (16.53,          - 2, -35.45), \nonumber\\
	\text{Point 7} &= (03.27,          - 2, -52.83), \quad
	\text{Point 8}  = (12.24,          - 2, -33.63).
\end{align} 
and
\begin{align}
	\text{Point 1'} &= (16.53, \phantom{-}1, -35.45), \quad
	\text{Point 2'}  = (19.57, \phantom{-}1, -29.63), \nonumber\\ 
	\text{Point 3'} &= (12.24, \phantom{-}1, -33.63), \quad
	\text{Point 4'}  = (14.57, \phantom{-}1, -28.63), \nonumber\\
	\text{Point 5'} &= (16.53,          - 2, -35.45), \quad
	\text{Point 6'}  = (19.57,          - 2, -29.63), \nonumber\\
	\text{Point 7'} &= (12.24,          - 2, -33.63), \quad
	\text{Point 8'}  = (14.57,          - 2, -28.63).
\end{align} 
With its larger volume and the increased luminosity, MAPP-2 is expected to capture ${\cal O}$(30) times the number of events of MAPP-1.

\subsection{Sensitivity to active-sterile neutrino mixing}
\label{sec:simu}	

\begin{figure}[t!]
	\centering
	\includegraphics[width=0.99\textwidth]{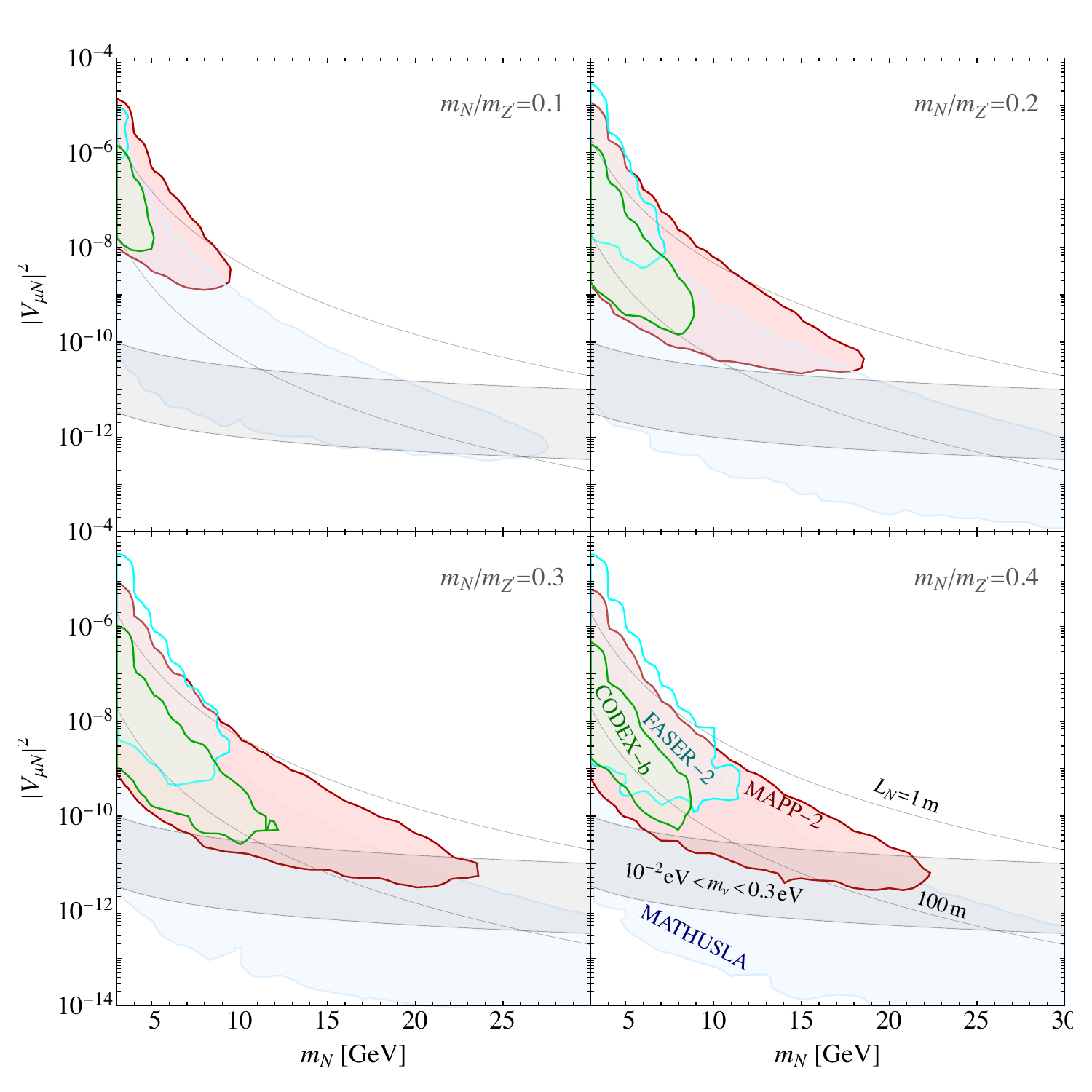}
	\caption{Projected sensitivity at 95\% confidence level of the MAPP-2 detector on the active-sterile mixing strength $V^2_{\mu N}$ as a function of the RH neutrino mass $m_N$ in the process $pp\to Z' \to NN$ for a luminosity of 300~fb$^{-1}$. Also shown are the corresponding sensitivities of the CODEX-b (300~fb$^{-1}$), FASER-2 and MATHUSLA (3000~fb$^{-1}$) detectors. The $U(1)_{B-L}$ gauge coupling and the $Z'$ mass are chosen as $g_{B-L} = 10^{-3}$ and $m_N/m_{Z'} = 0.1$ (top left), 0.2 (top right), 0.3 (bottom left) and 0.4 (bottom right). The grey curves are contours of constant proper RH neutrino decay lengths $L_N = 1$~m and 100~m. The grey shaded band indicates the preferred parameter region where the light neutrinos acquire a mass between $10^{-2}$~eV and $0.3$~eV in a canonical seesaw mechanism.}
	\label{fig:big}
\end{figure}
We first discuss the case of RH neutrino production via $Z'$ and its associated $B-L$ gauge coupling $g_{B-L}$. In Fig.~\ref{fig:big}, we show the sensitivity of MAPP-2 on the active-sterile neutrino mixing $V_{\mu N}^2$ as a function of $m_N$. As the MAPP-2 detector is far away from the interaction point and is protected by rock shielding and overburden, we assume that there is no background. We thus follow a Poisson distribution by highlighting the parameter space where there are more than 3.09 expected signal events at 95\% confidence level~\cite{ParticleDataGroup:2018ovx}. The $U(1)_{B-L}$ gauge coupling is $g_{B-L} = 10^{-3}$ and the $m_{Z'}$ is chosen such that the ratio $m_N / m_{Z'}$ is kept constant in each panel with values 0.1, 0.2, 0.3 and 0.4, as indicated. As the kinematic limit for on-shell $Z'$ decays into a pair of RH neutrino is $m_{Z'} > 2 m_N$, the sensitivity will decrease for larger values $m_N / m_{Z'} \to 0.5$. However, the production cross section for $Z'$ increases as $m_N / m_{Z'}$ goes up, since $Z'$ becomes lighter. Therefore, the sensitivity peaks when $m_N / m_{Z'} \approx 0.3$, reaching $m_N \lesssim 25$~GeV and $V^2_{\mu N} \gtrsim 10^{-11}$, covering a fraction of the grey shaded region where the Type-I seesaw mechanism is manifest. Overall, the sensitivity contours roughly track the region where $1~\text{m}< L_N < 100$~m.

Fig.~\ref{fig:big} also compares the sensitivity of MAPP-2 with that of CODEX-b, FASER-2 and MATHUSLA as representative examples of proposed long-lived particle detectors at the LHC. CODEX-b is planned to be built after the upcoming Run-3 upgrade of LHCb, and we use a corresponding luminosity of 300~fb$^{-1}$. It has a smaller angular acceptance and is located more forward than MAPP-2. Thus its sensitivity is weaker for the relatively heavy $Z'$ considered. Phase-1 of FASER is taking data at the LHC with 150~fb$^{-1}$ expected to be accumulated, but we do not find an appreciable sensitivity to our model. We instead show Phase-2 (FASER-2) proposed with an extended geometry for the 3000~fb$^{-1}$ HL-LHC. Although it would be operated with a larger integrated luminosity, its forward and narrow location requires the RH neutrinos to have a pseudo-rapidity of $\eta\approx 7$, therefore the sensitivity is weaker than MAPP-2. Finally, MATHUSLA is a proposed detector on the surface also designed for the HL-LHC with 3000~fb$^{-1}$. Due to its far distance located from interaction point, this ambitious proposal would have sensitivity to very small active-sterile mixing strengths as can be seen in Fig.~\ref{fig:big}. Long-lived particle searches at already existing LHC detectors typically probe larger active-sterile mixing strengths corresponding to shorter proper decay lengths $L_N \lesssim 1$~m and/or larger masses $m_N \gtrsim 25$~GeV, with a larger uncertainty due to potential background \cite{Deppisch:2019kvs}.

\begin{figure}[t!]
	\centering
	\includegraphics[width=0.99\textwidth]{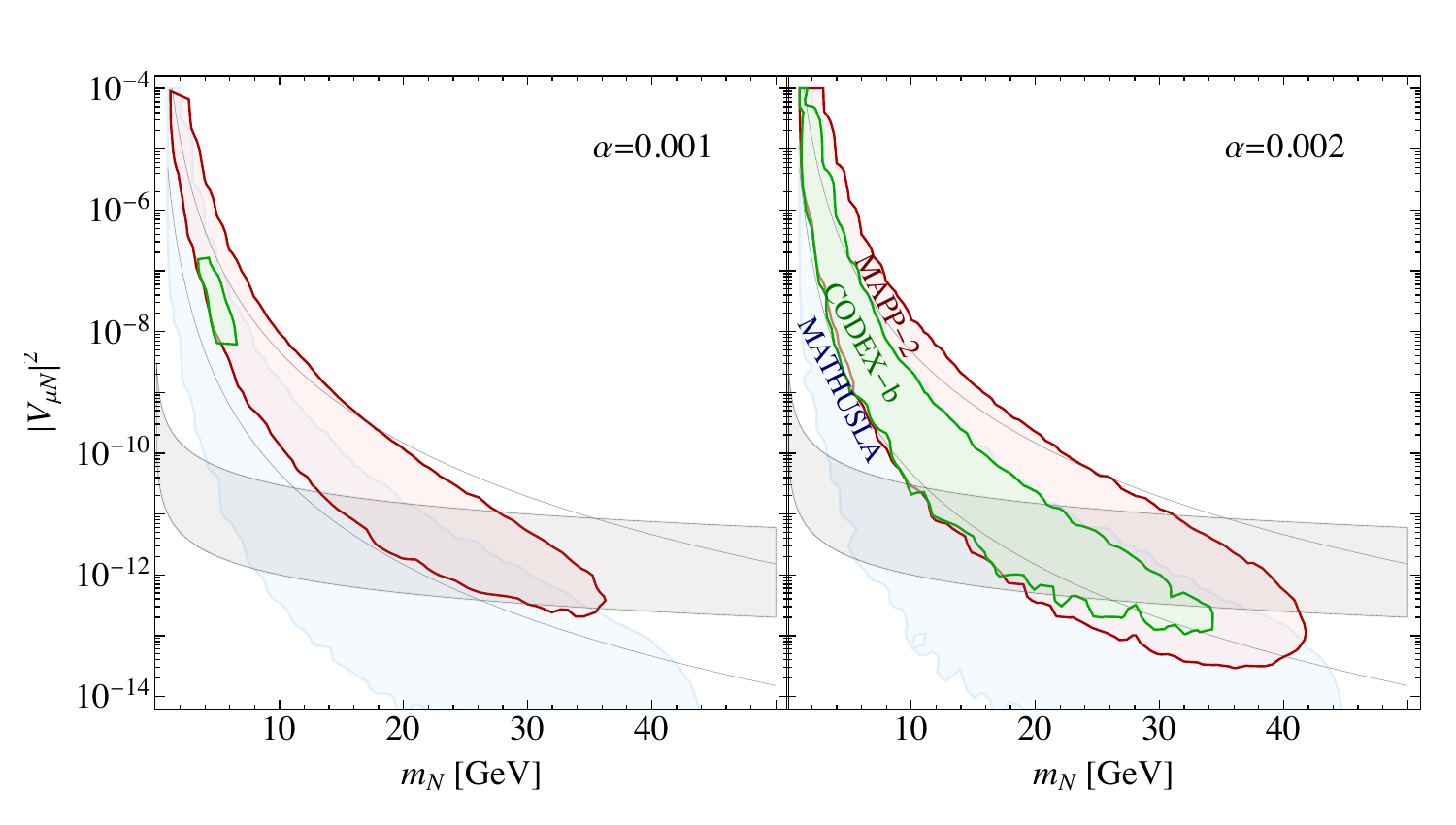}
	\caption{As Fig.~\ref{fig:big} but showing the projected sensitivity of MAPP-2 in the process $pp\to Z \to NN$. The gauge mixing parameter is $\alpha = g_{B-L}\sin\theta_{ZZ'} = 0.001$ (left) and $\alpha = 0.002$ (right).}
	\label{fig:gaugemixing}	
\end{figure}
Likewise, Fig.~\ref{fig:gaugemixing} shows the sensitivity of the process $pp \to Z\to NN$ via the gauge mixing parameter $\alpha = g_{B-L}\sin\theta_{ZZ'}$, see Eq.~\eqref{eq:decaygaugemixing}. MAPP-2 has sensitivity when $\alpha \gtrsim 0.001$ in which case it can still probe part of the seesaw region. With larger $\alpha = 0.002$, a much broader part of this interesting region is probed, and the sensitivity reaches to $m_N \lesssim 40$~GeV and $V^2_{\mu N} \gtrsim 10^{-13}$. In comparison, CODEX-b can also probe the Seesaw band for $\alpha = 0.002$, benefiting from the larger $Z$ production cross section for $\alpha = 0.002$. For this value, the production is stronger than in the $Z'$ case leading to a better sensitivity in $V_\mu^2$ in both cases, but the bands quickly become thinner and start to vanish for $\alpha = 0.001$.  MATHUSLA remains to have a strong sensitivity for small active-sterile neutrino mixing strengths whereas FASER-2 has no appreciable sensitivity.

\section{Conclusion}
\label{conclu}
In this work, we have discussed the potential to search for RH neutrinos at the MAPP detector facility. Within the framework of the $U(1)_{B-L}$ model, the RH neutrinos are pair-produced, mediated by either the $B-L$ gauge boson $Z'$ or the SM $Z$ boson. We have performed an analysis of searching for displaced RH neutrinos from such processes at the MAPP-2 detectors, requiring the RH neutrinos to decay within the decay volume of the detector, namely the MAPP-1 design at the LHC Run-3, and the proposed MAPP-2 design at the HL-LHC. The MAPP-1 detector is too small to have an appreciable sensitivity in our model. Since the MAPP-2 detector is placed roughly 50~meters away from the interaction point and it is protected by rock shielding, the background events are considered to be negligible. We obtain a sensitivity to probe RH neutrino with masses around $m_N \lesssim 20-30$~GeV. For the $Z'$ mediated channel, taking the $B-L$ gauge coupling $g_{B-L}$ at its current experimental limits, i.e., $g_{B-L} \approx 10^{-3}$, we show that the MAPP-2 detector can probe the interesting seesaw region of light neutrino mass generation where the active-sterile mixing is $V^2_{\mu N} \approx 10^{-12}$ in such a case if the $Z'$ boson mass is $m_{Z'} \approx (2.5 - 3.5)\times m_N$. The sensitivity of MAPP-2 is interesting as it is located between that of other displaced vertex searches \cite{Acharya:2022nik} and can specifically probe the seesaw regime for such RH neutrinos. We here compare the sensitivity of MAPP-2 with that of the similar proposed detectors CODEX-b, FASER-2 and MATHUSLA. When it comes to the SM $Z$ mediated channel, the cross section is controlled by the gauge mixing parameter $\alpha = g_{B-L}\sin\theta_{ZZ'}$ and the seesaw regime is probed as long as $\alpha \gtrsim 0.001$.

\begin{figure}[t!]
	\centering
	\includegraphics[width=0.99\textwidth]{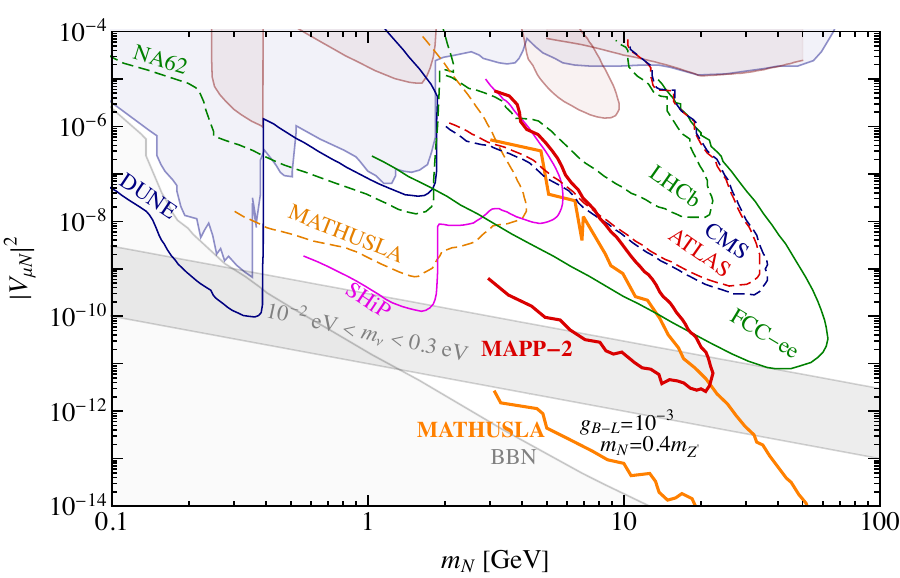}
	\caption{Current constraints (shaded regions) and future sensitivity on the active-sterile neutrino mixing strength $V^2_{\mu N}$ as a function of the sterile neutrino mass $m_N$. Here, the sterile neutrinos only couple through the mixing and they are produced via SM neutral or charged current, i.e., effectively, $g_{B-L} = \alpha = 0$. The corresponding data is taken from \href{sterile-neutrino.org}{http://sterile-neutrino.org}. In comparison, the future sensitivity of MAPP-2 and MATHUSLA in the $B-L$ model for fixed $g_{B-L} = 10^{-3}$ and $m_{N} = 0.4 m_{Z^\prime}$ is shown. The grey shaded region labelled BBN is disfavoured by Big Bang Nucleosynthesis whereas the grey shaded band indicates the preferred parameter region where the light neutrinos acquire a mass between $10^{-2}$~eV and $0.3$~eV in a canonical seesaw mechanism.}
	\label{fig:sterileconstraints}	
\end{figure}
The promising sensitivity to probe the origin of light neutrino mass generation in the $U(1)_{B-L}$ model relies on the efficient production of RH neutrinos via the exotic gauge interaction or mixing. In the sterile neutrino scenario, i.e., in absence of an exotic gauge interaction under which the RH neutrinos are charged, they are only produced via the same active-sterile mixing strength $V^2_{\mu N}$ and coupling with the SM $Z$ and $W$. This mixing is very small in the canonical seesaw regime, as required by the very light eV-scale neutrino masses. This makes probing the canonical seesaw regime very challenging. Current constraints and future sensitivities in this case are shown in Fig.~\ref{fig:sterileconstraints} in comparison with the sensitivity in the $B-L$ model for fixed $g_{B-L} = 10^{-3}$ and $m_N = 0.4 m_{Z'}$ at MAPP-2 and MATHUSLA for illustration. Even ambitious projects such as the FCC-ee, acting as an abundant $Z$ factory will struggle to reach the seesaw regime, whereas the abundant production in the $B-L$ model allows probing a considerable part of the canonical Seesaw region. We emphasize that the $B-L$ model makes stronger assumptions on the exotic physics needed to be present, whereas the production via SM currents is generally expected for sterile neutrinos. We also note that the improved sensitivity applies to other experiments and not just MAPP-2 and MATHUSLA which are shown here as examples for the purpose of  comparison in Fig.~\ref{fig:sterileconstraints}; other long-lived particle detectors can similarly benefit from the increased production rate still allowed in the $B-L$ model. This is illustrated by comparing the sensitivity of MATHUSLA in the SM-current case (thin dashed orange curve) with that in the $B-L$ model (thick solid curve). Analogous increases in sensitivity apply, for example, to SHiP, LHCb, CMS, ATLAS \cite{Padhan:2022fak} and FCC-ee \cite{Liu:2022kid}. Our analysis thus motivates exploring extended scenarios beyond sterile neutrinos such as the $B-L$ model, demonstrating the ability of displaced vertex searches to probe the origin of light neutrino masses.
	
\section*{Acknowldedgments}
F.F.D. acknowledges support from the UK Science and Technology Facilities Council (STFC) via the Consolidated Grants ST/P00072X/1 and ST/T000880/1. S.K. is supported by Elise-Richter grant project number V592-N27.  W.L. is supported by the 2021 Jiangsu Shuangchuang (Mass Innovation and Entrepreneurship) Talent Program (JSSCBS20210213), and the National Natural Science Foundation of China (Grant No.12205153).

\section*{Data Availability}
All results in this work were derived using simulations as detailed in the text. No original data was created apart from the results shown in the text.

\bibliography{F-S-W}

@article{A:2023wup,
	archiveprefix = {arXiv},
	author = {A., ShivaSankar K. and Das, Arindam and Lambiase, Gaetano and Nomura, Takaaki and Orikasa, Yuta},
	eprint = {2308.14483},
	month = {8},
	primaryclass = {hep-ph},
	title = {{Probing chiral and flavored $Z^\prime$ from cosmic bursts through neutrino interactions}},
	year = {2023}}

@article{Poddar:2022svz,
	archiveprefix = {arXiv},
	author = {Poddar, Tanmay Kumar and Goswami, Srubabati and Mishra, Arvind Kumar},
	doi = {10.1140/epjc/s10052-023-11367-4},
	eprint = {2206.03485},
	journal = {Eur. Phys. J. C},
	number = {3},
	pages = {223},
	primaryclass = {hep-ph},
	title = {{Energizing gamma ray bursts via $Z^{\prime }$ mediated neutrino heating}},
	volume = {83},
	year = {2023}}

@article{Davidson:1978pm,
	author = {Davidson, Aharon},
	doi = {10.1103/PhysRevD.20.776},
	journal = {Phys. Rev. D},
	pages = {776},
	reportnumber = {SU-4213-129, COO-3533-129},
	title = {B-L as the fourth color},
	volume = {20},
	year = {1979}}

@article{Lindner:2018kjo,
	archiveprefix = {arXiv},
	author = {Lindner, Manfred and Queiroz, Farinaldo S. and Rodejohann, Werner and Xu, Xun-Jie},
	doi = {10.1007/JHEP05(2018)098},
	eprint = {1803.00060},
	journal = {JHEP},
	pages = {098},
	primaryclass = {hep-ph},
	reportnumber = {IIPDM-2018-02},
	title = {{Neutrino-electron scattering: general constraints on Z' and dark photon models}},
	volume = {05},
	year = {2018}}

@article{Dev:2021otb,
	archiveprefix = {arXiv},
	author = {Dev, P. S. Bhupal and Rodejohann, Werner and Xu, Xun-Jie and Zhang, Yongchao},
	doi = {10.1007/JHEP06(2021)039},
	eprint = {2103.09067},
	journal = {JHEP},
	pages = {039},
	primaryclass = {hep-ph},
	title = {{Searching for Z' bosons at the P2 experiment}},
	volume = {06},
	year = {2021}}

@article{Alwall:2014hca,
	archiveprefix = {arXiv},
	author = {Alwall, J. and Frederix, R. and Frixione, S. and Hirschi, V. and Maltoni, F. and Mattelaer, O. and Shao, H. -S. and Stelzer, T. and Torrielli, P. and Zaro, M.},
	doi = {10.1007/JHEP07(2014)079},
	eprint = {1405.0301},
	journal = {JHEP},
	pages = {079},
	primaryclass = {hep-ph},
	reportnumber = {CERN-PH-TH-2014-064, CP3-14-18, LPN14-066, MCNET-14-09, ZU-TH-14-14},
	slaccitation = {%%CITATION = ARXIV:1405.0301;%%},
	title = {{The automated computation of tree-level and next-to-leading order differential cross sections, and their matching to parton shower simulations}},
	volume = {07},
	year = {2014}}

@article{Deppisch:2018eth,
	archiveprefix = {arXiv},
	author = {Deppisch, Frank F. and Liu, Wei and Mitra, Manimala},
	doi = {10.1007/JHEP08(2018)181},
	eprint = {1804.04075},
	journal = {JHEP},
	pages = {181},
	primaryclass = {hep-ph},
	reportnumber = {IP/BBSR/2018-4, IP-BBSR-2018-4},
	slaccitation = {%%CITATION = ARXIV:1804.04075;%%},
	title = {{Long-lived Heavy Neutrinos from Higgs Decays}},
	volume = {08},
	year = {2018}}

@article{Ilten:2018crw,
	archiveprefix = {arXiv},
	author = {Ilten, Philip and Soreq, Yotam and Williams, Mike and Xue, Wei},
	doi = {10.1007/JHEP06(2018)004},
	eprint = {1801.04847},
	journal = {JHEP},
	pages = {004},
	primaryclass = {hep-ph},
	reportnumber = {MIT-CTP/4976, CERN-TH-2017-282, MIT-CTP-4976},
	slaccitation = {%%CITATION = ARXIV:1801.04847;%%},
	title = {{Serendipity in dark photon searches}},
	volume = {06},
	year = {2018}}

@article{Fujiwara:1984mp,
	author = {Fujiwara, Takanori and Kugo, Taichiro and Terao, Haruhiko and Uehara, Shozo and Yamawaki, Koichi},
	doi = {10.1143/PTP.73.926},
	journal = {Prog. Theor. Phys.},
	pages = {926},
	reportnumber = {KUNS-764},
	slaccitation = {%%CITATION = PTPKA,73,926;%%},
	title = {{Nonabelian Anomaly and Vector Mesons as Dynamical Gauge Bosons of Hidden Local Symmetries}},
	volume = {73},
	year = {1985}}

@article{Degrande:2011ua,
	archiveprefix = {arXiv},
	author = {Degrande, Celine and Duhr, Claude and Fuks, Benjamin and Grellscheid, David and Mattelaer, Olivier and Reiter, Thomas},
	doi = {10.1016/j.cpc.2012.01.022},
	eprint = {1108.2040},
	journal = {Comput. Phys. Commun.},
	pages = {1201-1214},
	primaryclass = {hep-ph},
	reportnumber = {CP3-11-25, IPHC-PHENO-11-04, IPPP-11-39, DCPT-11-78, MPP-2011-68},
	slaccitation = {%%CITATION = ARXIV:1108.2040;%%},
	title = {{UFO - The Universal FeynRules Output}},
	volume = {183},
	year = {2012}}

@article{Mohapatra:1980qe,
	author = {Mohapatra, Rabindra N. and Marshak, R. E.},
	doi = {10.1103/PhysRevLett.44.1644.2, 10.1103/PhysRevLett.44.1316},
	journal = {Phys. Rev. Lett.},
	note = {[Erratum: Phys. Rev. Lett.44,1643(1980)]},
	pages = {1316-1319},
	reportnumber = {VPI-HEP-80/1},
	slaccitation = {%%CITATION = PRLTA,44,1316;%%},
	title = {{Local B-L Symmetry of Electroweak Interactions, Majorana Neutrinos and Neutron Oscillations}},
	volume = {44},
	year = {1980}}

@article{Aaboud:2017buh,
	archiveprefix = {arXiv},
	author = {Aaboud, Morad and others},
	collaboration = {ATLAS},
	doi = {10.1007/JHEP10(2017)182},
	eprint = {1707.02424},
	journal = {JHEP},
	pages = {182},
	primaryclass = {hep-ex},
	reportnumber = {CERN-EP-2017-119},
	slaccitation = {%%CITATION = ARXIV:1707.02424;%%},
	title = {{Search for new high-mass phenomena in the dilepton final state using 36 fb$^{-1}$ of proton-proton collision data at $ \sqrt{s}=13 $ TeV with the ATLAS detector}},
	volume = {10},
	year = {2017}}

@article{Heeck:2014zfa,
	archiveprefix = {arXiv},
	author = {Heeck, Julian},
	doi = {10.1016/j.physletb.2014.10.067},
	eprint = {1408.6845},
	journal = {Phys. Lett.},
	pages = {256-262},
	primaryclass = {hep-ph},
	slaccitation = {%%CITATION = ARXIV:1408.6845;%%},
	title = {{Unbroken B $-$ L symmetry}},
	volume = {B739},
	year = {2014}}

@article{Cacciapaglia:2006pk,
	archiveprefix = {arXiv},
	author = {Cacciapaglia, G. and Csaki, C. and Marandella, G. and Strumia, A.},
	doi = {10.1103/PhysRevD.74.033011},
	eprint = {hep-ph/0604111},
	journal = {Phys. Rev.},
	pages = {033011},
	primaryclass = {hep-ph},
	slaccitation = {%%CITATION = HEP-PH/0604111;%%},
	title = {{The Minimal Set of Electroweak Precision Parameters}},
	volume = {D74},
	year = {2006}}

@article{Anthony:2003ub,
	archiveprefix = {arXiv},
	author = {Anthony, P. L. and others},
	collaboration = {SLAC E158},
	doi = {10.1103/PhysRevLett.92.181602},
	eprint = {hep-ex/0312035},
	journal = {Phys. Rev. Lett.},
	pages = {181602},
	primaryclass = {hep-ex},
	reportnumber = {SLAC-PUB-10270},
	slaccitation = {%%CITATION = HEP-EX/0312035;%%},
	title = {{Observation of parity nonconservation in Moller scattering}},
	volume = {92},
	year = {2004}}

@article{LEP:2003aa,
	archiveprefix = {arXiv},
	author = {Electroweak, the SLD},
	collaboration = {SLD Electroweak Group, SLD Heavy Flavor Group, DELPHI, LEP, ALEPH, OPAL, LEP Electroweak Working Group, L3},
	eprint = {hep-ex/0312023},
	primaryclass = {hep-ex},
	reportnumber = {SLAC-R-701, LEPEWWG-2003-02, ALEPH-2003-017-PHYSICS-2003-005, DELPHI-2003-072-PHYS-937, L3-NOTE-2825, OPAL-PR-392, SLD-PHYSICS-NOTE-78, CERN-EP-2003-091, LEP-EWWG-2003-02},
	slaccitation = {%%CITATION = HEP-EX/0312023;%%},
	title = {{A Combination of preliminary electroweak measurements and constraints on the standard model}},
	year = {2003}}

@article{Carena:2004xs,
	archiveprefix = {arXiv},
	author = {Carena, Marcela and Daleo, Alejandro and Dobrescu, Bogdan A. and Tait, Timothy M. P.},
	doi = {10.1103/PhysRevD.70.093009},
	eprint = {hep-ph/0408098},
	journal = {Phys. Rev.},
	pages = {093009},
	primaryclass = {hep-ph},
	reportnumber = {FERMILAB-PUB-04-129-T},
	slaccitation = {%%CITATION = HEP-PH/0408098;%%},
	title = {{$Z^\prime$ gauge bosons at the Tevatron}},
	volume = {D70},
	year = {2004}}

@article{Accomando:2017qcs,
	archiveprefix = {arXiv},
	author = {Accomando, Elena and Delle Rose, Luigi and Moretti, Stefano and Olaiya, Emmanuel and Shepherd-Themistocleous, Claire H.},
	doi = {10.1007/JHEP02(2018)109},
	eprint = {1708.03650},
	journal = {JHEP},
	pages = {109},
	primaryclass = {hep-ph},
	slaccitation = {%%CITATION = ARXIV:1708.03650;%%},
	title = {{Extra Higgs boson and Z' as portals to signatures of heavy neutrinos at the LHC}},
	volume = {02},
	year = {2018}}

@article{Antusch:2017hhu,
	archiveprefix = {arXiv},
	author = {Antusch, Stefan and Cazzato, Eros and Fischer, Oliver},
	doi = {10.1016/j.physletb.2017.09.057},
	eprint = {1706.05990},
	journal = {Phys. Lett.},
	pages = {114-118},
	primaryclass = {hep-ph},
	slaccitation = {%%CITATION = ARXIV:1706.05990;%%},
	title = {{Sterile neutrino searches via displaced vertices at LHCb}},
	volume = {B774},
	year = {2017}}

@article{Ramirez-Sanchez:2016ugz,
	archiveprefix = {arXiv},
	author = {Ramirez-Sanchez, F. and Gutierrez-Rodriguez, A. and Hernandez-Ruiz, M. A.},
	doi = {10.1088/0954-3899/43/9/095003},
	eprint = {1606.04144},
	journal = {J. Phys.},
	number = {9},
	pages = {095003},
	primaryclass = {hep-ph},
	slaccitation = {%%CITATION = ARXIV:1606.04144;%%},
	title = {{Higgs bosons production and decay at future $e^+e^-$ linear colliders as a probe of the B $-$ L model}},
	volume = {G43},
	year = {2016}}

@article{Amrith:2018yfb,
	archiveprefix = {arXiv},
	author = {Amrith, S. and Butterworth, J. M. and Deppisch, F. F. and Liu, W. and Varma, A. and Yallup, D.},
	eprint = {1811.11452},
	primaryclass = {hep-ph},
	reportnumber = {MCnet-18-30},
	slaccitation = {%%CITATION = ARXIV:1811.11452;%%},
	title = {{LHC Constraints on a $B-L$ Gauge Model using Contur}},
	year = {2018}}

@article{Butterworth:2016sqg,
	archiveprefix = {arXiv},
	author = {Butterworth, Jonathan M. and Grellscheid, David and Kr$\ddot{\text{a}}$mer, Michael and Sarrazin, Bj$\ddot{\text{o}}$rn and Yallup, David},
	doi = {10.1007/JHEP03(2017)078},
	eprint = {1606.05296},
	journal = {JHEP},
	pages = {078},
	primaryclass = {hep-ph},
	reportnumber = {IPPP-16-52, MCNET-16-21, TTK-16-22},
	slaccitation = {%%CITATION = ARXIV:1606.05296;%%},
	title = {{Constraining new physics with collider measurements of Standard Model signatures}},
	volume = {03},
	year = {2017}}

@article{Ariga:2018uku,
	archiveprefix = {arXiv},
	author = {Ariga, Akitaka and others},
	collaboration = {FASER},
	doi = {10.1103/PhysRevD.99.095011},
	eprint = {1811.12522},
	journal = {Phys. Rev.},
	number = {9},
	pages = {095011},
	primaryclass = {hep-ph},
	reportnumber = {UCI-TR-2018-19, KYUSHU-RCAPP-2018-06},
	slaccitation = {%%CITATION = ARXIV:1811.12522;%%},
	title = {{FASER's Physics Reach for Long-Lived Particles}},
	volume = {D99},
	year = {2019}}

@article{Batell:2016zod,
	archiveprefix = {arXiv},
	author = {Batell, Brian and Pospelov, Maxim and Shuve, Brian},
	doi = {10.1007/JHEP08(2016)052},
	eprint = {1604.06099},
	journal = {JHEP},
	pages = {052},
	primaryclass = {hep-ph},
	slaccitation = {%%CITATION = ARXIV:1604.06099;%%},
	title = {{Shedding Light on Neutrino Masses with Dark Forces}},
	volume = {08},
	year = {2016}}

@article{Alekhin:2015byh,
	archiveprefix = {arXiv},
	author = {Alekhin, Sergey and others},
	doi = {10.1088/0034-4885/79/12/124201},
	eprint = {1504.04855},
	journal = {Rept. Prog. Phys.},
	number = {12},
	pages = {124201},
	primaryclass = {hep-ph},
	reportnumber = {CERN-SPSC-2015-017, SPSC-P-350-ADD-1},
	slaccitation = {%%CITATION = ARXIV:1504.04855;%%},
	title = {{A facility to Search for Hidden Particles at the CERN SPS: the SHiP physics case}},
	volume = {79},
	year = {2016}}

@article{Anelli:2015pba,
	archiveprefix = {arXiv},
	author = {Anelli, M. and others},
	collaboration = {SHiP},
	eprint = {1504.04956},
	primaryclass = {physics.ins-det},
	reportnumber = {CERN-SPSC-2015-016, SPSC-P-350},
	slaccitation = {%%CITATION = ARXIV:1504.04956;%%},
	title = {{A facility to Search for Hidden Particles (SHiP) at the CERN SPS}},
	year = {2015}}

@article{CMS:2014hka,
	archiveprefix = {arXiv},
	author = {Khachatryan, Vardan and others},
	collaboration = {CMS},
	doi = {10.1103/PhysRevD.91.052012},
	eprint = {1411.6977},
	journal = {Phys. Rev.},
	number = {5},
	pages = {052012},
	primaryclass = {hep-ex},
	reportnumber = {CMS-EXO-12-037, CERN-PH-EP-2014-263},
	slaccitation = {%%CITATION = ARXIV:1411.6977;%%},
	title = {{Search for long-lived particles that decay into final states containing two electrons or two muons in proton-proton collisions at $\sqrt{s} =$ 8 TeV}},
	volume = {D91},
	year = {2015}}

@article{CMS:2015pca,
	author = {CMS Collaboration},
	collaboration = {CMS},
	reportnumber = {CMS-PAS-EXO-14-012},
	slaccitation = {%%CITATION = CMS-PAS-EXO-14-012;%%},
	title = {{Search for long-lived particles that decay into final states containing two muons, reconstructed using only the CMS muon chambers}},
	year = {2015}}

@article{Pinfold:2019nqj,
	author = {Pinfold, James Lewis},
	booktitle = {{Proceedings, 7th International Conference on New Frontiers in Physics (ICNFP 2018): Kolymbari, Crete, Greece, July 4-12, 2018}},
	doi = {10.3390/universe5020047},
	journal = {Universe},
	number = {2},
	pages = {47},
	slaccitation = {%%CITATION = INSPIRE-1719295;%%},
	title = {{The MoEDAL Experiment at the LHC-A Progress Report}},
	volume = {5},
	year = {2019}}

@article{Chou:2016lxi,
	archiveprefix = {arXiv},
	author = {Chou, John Paul and Curtin, David and Lubatti, H. J.},
	doi = {10.1016/j.physletb.2017.01.043},
	eprint = {1606.06298},
	journal = {Phys. Lett.},
	pages = {29-36},
	primaryclass = {hep-ph},
	slaccitation = {%%CITATION = ARXIV:1606.06298;%%},
	title = {{New Detectors to Explore the Lifetime Frontier}},
	volume = {B767},
	year = {2017}}

@article{Gligorov:2017nwh,
	archiveprefix = {arXiv},
	author = {Gligorov, Vladimir V. and Knapen, Simon and Papucci, Michele and Robinson, Dean J.},
	doi = {10.1103/PhysRevD.97.015023},
	eprint = {1708.09395},
	journal = {Phys. Rev.},
	number = {1},
	pages = {015023},
	primaryclass = {hep-ph},
	slaccitation = {%%CITATION = ARXIV:1708.09395;%%},
	title = {{Searching for Long-lived Particles: A Compact Detector for Exotics at LHCb}},
	volume = {D97},
	year = {2018}}

@article{Aad:2019kiz,
	archiveprefix = {arXiv},
	author = {Aad, Georges and others},
	collaboration = {ATLAS},
	eprint = {1905.09787},
	primaryclass = {hep-ex},
	reportnumber = {CERN-EP-2019-071},
	slaccitation = {%%CITATION = ARXIV:1905.09787;%%},
	title = {{Search for heavy neutral leptons in decays of $W$ bosons produced in 13 TeV $pp$ collisions using prompt and displaced signatures with the ATLAS detector}},
	year = {2019}}

@article{Sirunyan:2018mtv,
	archiveprefix = {arXiv},
	author = {Sirunyan, Albert M and others},
	collaboration = {CMS},
	doi = {10.1103/PhysRevLett.120.221801},
	eprint = {1802.02965},
	journal = {Phys. Rev. Lett.},
	number = {22},
	pages = {221801},
	primaryclass = {hep-ex},
	reportnumber = {CMS-EXO-17-012, CERN-EP-2018-006},
	slaccitation = {%%CITATION = ARXIV:1802.02965;%%},
	title = {{Search for heavy neutral leptons in events with three charged leptons in proton-proton collisions at $\sqrt{s} =$ 13 TeV}},
	volume = {120},
	year = {2018}}

@article{CMS:2014mca,
	author = {CMS Collaboration},
	collaboration = {CMS},
	reportnumber = {CMS-PAS-EXO-12-037},
	slaccitation = {%%CITATION = CMS-PAS-EXO-12-037;%%},
	title = {{Search for long-lived particles decaying to final states that include dileptons}},
	year = {2014}}

@article{Khachatryan:2015gha,
	archiveprefix = {arXiv},
	author = {Khachatryan, Vardan and others},
	collaboration = {CMS},
	doi = {10.1016/j.physletb.2015.06.070},
	eprint = {1501.05566},
	journal = {Phys. Lett.},
	pages = {144-166},
	primaryclass = {hep-ex},
	reportnumber = {CMS-EXO-12-057, CERN-PH-EP-2015-001},
	slaccitation = {%%CITATION = ARXIV:1501.05566;%%},
	title = {{Search for heavy Majorana neutrinos in $\mu^\pm \mu^\pm+$ jets events in proton-proton collisions at $\sqrt{s}$ = 8 TeV}},
	volume = {B748},
	year = {2015}}

@article{Ade:2015xua,
	archiveprefix = {arXiv},
	author = {Ade, P. A. R. and others},
	collaboration = {Planck},
	doi = {10.1051/0004-6361/201525830},
	eprint = {1502.01589},
	journal = {Astron. Astrophys.},
	pages = {A13},
	primaryclass = {astro-ph.CO},
	slaccitation = {%%CITATION = ARXIV:1502.01589;%%},
	title = {{Planck 2015 results. XIII. Cosmological parameters}},
	volume = {594},
	year = {2016}}

@article{Deppisch:2015qwa,
	archiveprefix = {arXiv},
	author = {Deppisch, Frank F. and Bhupal Dev, P. S. and Pilaftsis, Apostolos},
	doi = {10.1088/1367-2630/17/7/075019},
	eprint = {1502.06541},
	journal = {New J. Phys.},
	number = {7},
	pages = {075019},
	primaryclass = {hep-ph},
	reportnumber = {MAN-HEP-2014-15},
	slaccitation = {%%CITATION = ARXIV:1502.06541;%%},
	title = {{Neutrinos and Collider Physics}},
	volume = {17},
	year = {2015}}

@article{Kraus:2004zw,
	archiveprefix = {arXiv},
	author = {Kraus, Ch. and others},
	doi = {10.1140/epjc/s2005-02139-7},
	eprint = {hep-ex/0412056},
	journal = {Eur. Phys. J.},
	pages = {447-468},
	primaryclass = {hep-ex},
	slaccitation = {%%CITATION = HEP-EX/0412056;%%},
	title = {{Final results from phase II of the Mainz neutrino mass search in tritium beta decay}},
	volume = {C40},
	year = {2005}}

@article{Aseev:2011dq,
	archiveprefix = {arXiv},
	author = {Aseev, V. N. and others},
	collaboration = {Troitsk},
	doi = {10.1103/PhysRevD.84.112003},
	eprint = {1108.5034},
	journal = {Phys. Rev.},
	pages = {112003},
	primaryclass = {hep-ex},
	slaccitation = {%%CITATION = ARXIV:1108.5034;%%},
	title = {{An upper limit on electron antineutrino mass from Troitsk experiment}},
	volume = {D84},
	year = {2011}}

@article{Sjostrand:2014zea,
	archiveprefix = {arXiv},
	author = {Sj$\ddot{\text{o}}$strand, Torbj$\ddot{\text{o}}$rn and Ask, Stefan and Christiansen, Jesper R. and Corke, Richard and Desai, Nishita and Ilten, Philip and Mrenna, Stephen and Prestel, Stefan and Rasmussen, Christine O. and Skands, Peter Z.},
	doi = {10.1016/j.cpc.2015.01.024},
	eprint = {1410.3012},
	journal = {Comput. Phys. Commun.},
	pages = {159-177},
	primaryclass = {hep-ph},
	reportnumber = {LU-TP-14-36, MCNET-14-22, CERN-PH-TH-2014-190, FERMILAB-PUB-14-316-CD, DESY-14-178, SLAC-PUB-16122},
	slaccitation = {%%CITATION = ARXIV:1410.3012;%%},
	title = {{An Introduction to PYTHIA 8.2}},
	volume = {191},
	year = {2015}}

@article{Das:2017flq,
	archiveprefix = {arXiv},
	author = {Das, Arindam and Okada, Nobuchika and Raut, Digesh},
	doi = {10.1103/PhysRevD.97.115023},
	eprint = {1710.03377},
	journal = {Phys. Rev.},
	number = {11},
	pages = {115023},
	primaryclass = {hep-ph},
	slaccitation = {%%CITATION = ARXIV:1710.03377;%%},
	title = {{Enhanced pair production of heavy Majorana neutrinos at the LHC}},
	volume = {D97},
	year = {2018}}

@article{Das:2017deo,
	archiveprefix = {arXiv},
	author = {Das, Arindam and Okada, Nobuchika and Raut, Digesh},
	doi = {10.1140/epjc/s10052-018-6171-8},
	eprint = {1711.09896},
	journal = {Eur. Phys. J.},
	number = {9},
	pages = {696},
	primaryclass = {hep-ph},
	slaccitation = {%%CITATION = ARXIV:1711.09896;%%},
	title = {{Heavy Majorana neutrino pair productions at the LHC in minimal U(1) extended Standard Model}},
	volume = {C78},
	year = {2018}}

@article{Das:2018tbd,
	archiveprefix = {arXiv},
	author = {Das, Arindam and Okada, Nobuchika and Okada, Satomi and Raut, Digesh},
	eprint = {1812.11931},
	primaryclass = {hep-ph},
	reportnumber = {OU-HEP-994},
	slaccitation = {%%CITATION = ARXIV:1812.11931;%%},
	title = {{Probing the seesaw mechanism at the 250 GeV ILC}},
	year = {2018}}

@article{Jana:2018rdf,
	archiveprefix = {arXiv},
	author = {Jana, Sudip and Okada, Nobuchika and Raut, Digesh},
	doi = {10.1103/PhysRevD.98.035023},
	eprint = {1804.06828},
	journal = {Phys. Rev.},
	number = {3},
	pages = {035023},
	primaryclass = {hep-ph},
	reportnumber = {OSU-HEP-18-02, FERMILAB-PUB-18-176-T},
	slaccitation = {%%CITATION = ARXIV:1804.06828;%%},
	title = {{Displaced vertex signature of type-I seesaw model}},
	volume = {D98},
	year = {2018}}

@article{Khalil:2006yi,
	archiveprefix = {arXiv},
	author = {Khalil, Shaaban},
	doi = {10.1088/0954-3899/35/5/055001},
	eprint = {hep-ph/0611205},
	journal = {J. Phys.},
	pages = {055001},
	primaryclass = {hep-ph},
	slaccitation = {%%CITATION = HEP-PH/0611205;%%},
	title = {{Low scale $B$ - L extension of the Standard Model at the LHC}},
	volume = {G35},
	year = {2008}}

@article{Khalil:2010iu,
	archiveprefix = {arXiv},
	author = {Khalil, Shaaban},
	doi = {10.1103/PhysRevD.82.077702},
	eprint = {1004.0013},
	journal = {Phys. Rev.},
	pages = {077702},
	primaryclass = {hep-ph},
	slaccitation = {%%CITATION = ARXIV:1004.0013;%%},
	title = {{TeV-scale gauged B-L symmetry with inverse seesaw mechanism}},
	volume = {D82},
	year = {2010}}

@article{Dib:2014fua,
	archiveprefix = {arXiv},
	author = {Dib, Claudio O. and Moreno, Gaston R. and Neill, Nicolas A.},
	doi = {10.1103/PhysRevD.90.113003},
	eprint = {1409.1868},
	journal = {Phys. Rev.},
	number = {11},
	pages = {113003},
	primaryclass = {hep-ph},
	slaccitation = {%%CITATION = ARXIV:1409.1868;%%},
	title = {{Neutrinos with a linear seesaw mechanism in a scenario of gauged B-L symmetry}},
	volume = {D90},
	year = {2014}}

@article{ParticleDataGroup:2018ovx,
	author = {Tanabashi, M. and others},
	collaboration = {Particle Data Group},
	doi = {10.1103/PhysRevD.98.030001},
	journal = {Phys. Rev. D},
	number = {3},
	pages = {030001},
	title = {{Review of Particle Physics}},
	volume = {98},
	year = {2018}}

@article{ATLAS:2016fij,
	archiveprefix = {arXiv},
	author = {Aad, Georges and others},
	collaboration = {ATLAS},
	doi = {10.1016/j.physletb.2016.06.023},
	eprint = {1603.09222},
	journal = {Phys. Lett. B},
	pages = {601--621},
	primaryclass = {hep-ex},
	reportnumber = {CERN-EP-2016-069},
	title = {{Measurement of $W^{\pm}$ and $Z$-boson production cross sections in $pp$ collisions at $\sqrt{s}=13$ TeV with the ATLAS detector}},
	volume = {759},
	year = {2016}}

@article{Ogul:2017zjd,
	archiveprefix = {arXiv},
	author = {Ogul, Hasan and Dilsiz, Kamuran},
	doi = {10.1155/2017/8262018},
	eprint = {1702.07206},
	journal = {Adv. High Energy Phys.},
	pages = {8262018},
	primaryclass = {hep-ph},
	title = {{Cross Section Prediction for Inclusive Production of Z Boson in $pp$ Collisions at $\sqrt{s}=14$ TeV: A Study of Systematic Uncertainty Due to Scale Dependence}},
	volume = {2017},
	year = {2017}}

@article{Bolton:2019pcu,
	archiveprefix = {arXiv},
	author = {Bolton, Patrick D. and Deppisch, Frank F. and Bhupal Dev, P. S.},
	doi = {10.1007/JHEP03(2020)170},
	eprint = {1912.03058},
	journal = {JHEP},
	pages = {170},
	primaryclass = {hep-ph},
	title = {{Neutrinoless double beta decay versus other probes of heavy sterile neutrinos}},
	volume = {03},
	year = {2020}}

@article{Bolton:2022pyf,
	archiveprefix = {arXiv},
	author = {Bolton, Patrick D. and Deppisch, Frank F. and Dev, P. S. Bhupal},
	booktitle = {{56th Rencontres de Moriond on Electroweak Interactions and Unified Theories}},
	eprint = {2206.01140},
	month = {6},
	primaryclass = {hep-ph},
	title = {{Probes of Heavy Sterile Neutrinos}},
	year = {2022}}

@article{Abdullahi:2022jlv,
	archiveprefix = {arXiv},
	author = {Abdullahi, Asli M. and others},
	booktitle = {{2022 Snowmass Summer Study}},
	eprint = {2203.08039},
	month = {3},
	primaryclass = {hep-ph},
	reportnumber = {FERMILAB-CONF-22-184-T-V},
	title = {{The Present and Future Status of Heavy Neutral Leptons}},
	year = {2022}}

@article{Feng:2022inv,
	archiveprefix = {arXiv},
	author = {Feng, Jonathan L. and others},
	eprint = {2203.05090},
	month = {3},
	primaryclass = {hep-ex},
	reportnumber = {UCI-TR-2022-01, CERN-PBC-Notes-2022-001, INT-PUB-22-006, BONN-TH-2022-04, FERMILAB-PUB-22-094-ND-SCD-T},
	title = {{The Forward Physics Facility at the High-Luminosity LHC}},
	year = {2022}}

@article{MammenAbraham:2020hex,
	author = {Mammen Abraham, Roshan and others},
	doi = {10.5281/zenodo.4059893},
	month = {4},
	title = {{Forward Physics Facility - Snowmass 2021 Letter of Interest}},
	year = {2020}}

@article{Alimena:2019zri,
	archiveprefix = {arXiv},
	author = {Alimena, Juliette and others},
	doi = {10.1088/1361-6471/ab4574},
	eprint = {1903.04497},
	journal = {J. Phys. G},
	number = {9},
	pages = {090501},
	primaryclass = {hep-ex},
	title = {{Searching for long-lived particles beyond the Standard Model at the Large Hadron Collider}},
	volume = {47},
	year = {2020}}

@article{Acharya:2022nik,
	archiveprefix = {arXiv},
	author = {Acharya, B. and others},
	booktitle = {{2022 Snowmass Summer Study}},
	eprint = {2209.03988},
	month = {9},
	primaryclass = {hep-ph},
	title = {{MoEDAL-MAPP, an LHC Dedicated Detector Search Facility}},
	year = {2022}}

@article{Deppisch:2019kvs,
	archiveprefix = {arXiv},
	author = {Deppisch, Frank and Kulkarni, Suchita and Liu, Wei},
	doi = {10.1103/PhysRevD.100.035005},
	eprint = {1905.11889},
	journal = {Phys. Rev. D},
	number = {3},
	pages = {035005},
	primaryclass = {hep-ph},
	title = {{Heavy neutrino production via $Z'$ at the lifetime frontier}},
	volume = {100},
	year = {2019}}

@article{Liu:2022kid,
	archiveprefix = {arXiv},
	author = {Liu, Wei and Kulkarni, Suchita and Deppisch, Frank F.},
	doi = {10.1103/PhysRevD.105.095043},
	eprint = {2202.07310},
	journal = {Phys. Rev. D},
	number = {9},
	pages = {095043},
	primaryclass = {hep-ph},
	title = {{Heavy neutrinos at the FCC-hh in the U(1)B-L model}},
	volume = {105},
	year = {2022}}

@article{Padhan:2022fak,
	archiveprefix = {arXiv},
	author = {Padhan, Rojalin and Mitra, Manimala and Kulkarni, Suchita and Deppisch, Frank F.},
	doi = {10.1140/epjc/s10052-022-10819-7},
	eprint = {2203.06114},
	journal = {Eur. Phys. J. C},
	number = {10},
	pages = {858},
	primaryclass = {hep-ph},
	title = {{Displaced fat-jets and tracks to probe boosted right-handed neutrinos in the $U(1)_{B-L}$ model}},
	volume = {82},
	year = {2022}}

@article{Deppisch:2016rox,
	archiveprefix = {arXiv},
	author = {Deppisch, Frank F. and Suhonen, Jouni},
	doi = {10.1103/PhysRevC.94.055501},
	eprint = {1606.02908},
	journal = {Phys. Rev. C},
	number = {5},
	pages = {055501},
	primaryclass = {nucl-th},
	title = {{Statistical analysis of $\beta$ decays and the effective value of $g_A$ in the proton-neutron quasiparticle random-phase approximation framework}},
	volume = {94},
	year = {2016}}
\end{document}